\documentclass[nofootinbib,
 reprint, 
 amsmath,amssymb,
 aps,
]{revtex4-2}

\newcommand{\ca}{\mathcal A}
\newcommand{\cb}{\mathcal B}
\newcommand{\cc}{\mathcal C}
\newcommand{\cd}{\mathcal D}

\newcommand{\ch}{\mathcal H}

\newcommand{\cs}{\mathcal S}

\newcommand{\cu}{\mathcal U}

\newcommand{\cz}{\mathcal Z}

\newtheorem{theorem}{Theorem}
\newtheorem{lemma}{Lemma}
\newtheorem{definition}{Definition}

\usepackage{graphicx, braket, amsfonts, amsmath, mathrsfs, xcolor, appendix, hyperref}%
\renewcommand{\url}[1]{\texttt{#1}}
\AtBeginDocument{%
  \renewcommand{\href}[2]{#2} 
}
\usepackage{dcolumn}
\usepackage{bm}

\begin{document}

\preprint{APS/123-QED}

\title{A causal derivation of the algebraic approach to quantum systems}

\author{Nick Ormrod}
\email{normrod@perimeterinstitute.ca.}
\affiliation{Perimeter Institute for Theoretical Physics, 31 Caroline Street North, Waterloo, Ontario Canada N2L 2Y5}

\begin{abstract} 

It is commonly assumed that every quantum system is represented by some algebra of operators. Doubt is cast on this assumption by what appears, at first glance, to be a reasonable candidate for a quantum system that is not naturally represented by any algebra. To resolve this puzzle, this work draws inspiration from recent frameworks for causal modelling in quantum theory to propose a ``causal view'' of quantum systems. The causal view defines quantum systems purely in terms of the causal structure of the unitary dynamics. The algebraic representation of quantum systems is derived from the causal view: it is proven that every quantum system corresponds to a unique von Neumann algebra of operators.
The causal view is extended with a definition of a ``classical quantum system'' inspired by quantum Darwinism. It is shown that such a system corresponds to a unique \textit{commutative} von Neumann operator algebra, completing the derivation of the traditional algebraic approach to quantum systems from the causal view.
The causal view is contrasted with the ``epistemic view'' of quantum systems, which is incompatible with the algebraic approach.
\end{abstract}
\maketitle

\section{Introduction} \label{sec:introduction}
Traditionally, it has often been assumed that the most general sort of quantum system is represented by an algebra of operators (most famously in algebraic quantum field theory \cite{haag1964algebraic}, more recently in quantum information and foundations \cite{viola2001constructing, zanardi2001virtual, zanardi2004quantum, ali2022quantum, vanrietvelde2025partitionsquantumtheory}), and that the algebra is commutative when the quantum system is ``classical''. We call this the \textit{algebraic approach to quantum systems}.
This paper asks how, or even whether, the algebraic approach can be justified.

The algebraic approach recovers the familiar notion of a system as a Hilbert space, since any Hilbert space $\ch$ can be identified with the algebra of bounded operators $\cb(\ch)$. But the algebraic approach is also more general. If a superselection rule forbids superpositions of states with respect to a preferred direct sum decomposition $\ch = \bigoplus_i \ch_i$ of the Hilbert space, then the superselected system can be identified with the algebra of block-diagonal bounded operators $\bigoplus_i \cb(\ch_i)$.

But is the algebraic approach general enough? To motivate this question, we note that there are certain sets of operators in quantum theory that are not operator algebras in their own right, do not naturally correspond to operator algebras, and yet do initially seem plausible as candidates for quantum systems. For example, consider the set of all real linear combinations of the Pauli observables $I$, $X$, and $Z$ on a qubit:
\begin{equation} \label{eq:rebit}
    \mathscr{R} = {\rm rspan}(I, X, Z).
\end{equation}
To see why one might reasonably suggest that $\mathscr{R}$ defines a subsystem of the qubit, suppose that an observer Polly performs many measurements of $X$ and $Z$ on an ensemble of identically prepared qubits in order to determine the expectation values $\langle X\rangle_\rho$ and $\langle Z\rangle_\rho$. From these expectation values she can immediately deduce the expectations of all observables in $\mathscr{R}$. She cannot in general deduce the expectation values of any observables outside of this set (she can in special cases, but see Appendix \ref{app:modifed}). Therefore, $\mathscr{R}$ is a \textit{full list of observables whose expectation values are known to some possible observer}. 

One might think that this last fact alone qualifies a set of observables to represent a quantum system. We call this the \textit{epistemic view of systems} (not to be confused with the logically unrelated epistemic view of quantum states defended in e.g.\ \cite{Spekkens_2007}). On the epistemic view, $\mathscr{R}$ represents a quantum system. 

Corroborating this conclusion, in the context of generalized probabilistic theories (GPTs) \cite{hardy2001quantum, barrett2007information}, \cite{schmid2024shadows} proposes a formal definition of a ``GPT subsystem'' according to which $\mathscr{R}$ is a GPT subsystem of the qubit.\footnote{However, the authors of \cite{schmid2024shadows} take care to note that the notion of a GPT system is necessarily highly general since arbitrary GPTs lack the algebraic structure that traditionally features in the definition of a quantum system. They do not take a position on whether all GPT subsystems in a formulation of quantum theory as a GPT are quantum subsystems: ``the rebit is typically not considered to be a \textit{quantum}
subsystem of a qubit, but it makes sense to consider it a
\textit{GPT} subsystem'' (original emphasis).} Furthermore, $\mathscr{R}$ defines the most elementary system (a ``rebit'') in a restricted version of quantum theory known as ``real quantum theory'' \cite{stueckelberg1960quantum, caves2001entanglement}.

However, $\mathscr{R}$ does not naturally correspond to any algebra. Formally, there is no (von Neumann or $C^*$-) algebra $\ca$ such that $\mathscr{R}$ is the full set of its observables, $\mathscr{R} = {\rm Herm}(\ca)$. An easy way to see this is to note that any algebra contains arbitrary complex rescalings and products of its members. Since $\mathscr{R}$ contains $X$ and $Z$, any algebra containing $\mathscr{R}$ must also contain $Y = iXZ$. However, $\mathscr{R}$ does not contain $Y$.

But should this prevent us from viewing $\mathscr{R}$ as a system? There is no obvious, direct operational meaning of the product of two observables (which in general is not an observable itself), so it is not obvious why it should be a problem that $X, Z \in \mathscr{R}$ but $Y \not\in \mathscr{R}$. To justify the algebraic approach to systems, one needs to convincingly explain what is ``wrong'' with viewing $\mathscr{R}$ as a system, which in part means explaining why closure under products is relevant for systemhood.

Rising to this challenge, this paper draws inspiration from the recent ``causal turn'' in the foundations of quantum physics. Among other things, the inadequacy of standard causal models \cite{pearl2009causality, spirtes2000causation} for explaining quantum correlations \cite{Wood_2015, cavalcanti2018classical, pearl2021classical} and the possibility of ``indefinite causal order'' \cite{Hardy:2010cm, Chiribella_2009, Chiribella_2013, oreshkov2012quantum} have led many to suspect that a radical concept of causation plays an essential role in quantum theory (see e.g.\ \cite{spekkens2015paradigm}). This view has motivated the development of intrinsically quantum frameworks for causal modelling \cite{pienaar2015graph, costa2016, Allen_2017, pienaar2019time, pienaar2020quantum, barrett2020quantum, barrett2021cyclic, ormrod2023causal}, and more recently an interpretation of the theory in which causation is more fundamental than states \cite{ormrod2024quantum}. This paper contributes a \textit{causal view of quantum systems}, formally defining a quantum system in terms of the causal structure of unitary interactions. Heuristically, the causal view tells us that a system can be thought of as a \textit{full list of observables that are accessible to some possible observer}.

Also inspired by quantum Darwinism \cite{zurek2003decoherence, zurek2009quantum}, the causal view defines a \textit{classical} quantum system as a full list of observables that are \textit{independently accessible} to a \textit{pair} of possible observers. Again, the intuition here is formally pinned down in terms of causal influences through unitary interactions.

This paper rigorously derives the algebraic approach to quantum systems from the causal view. Specifically, it is proven that any quantum system is represented by a unique von Neumann algebra $\ca$, which is commutative in the case of classical quantum systems. The result explains what is ``wrong'' with viewing $\mathscr{R}$ as a quantum system, and supports the idea that causation lies at the heart of quantum theory.

\section{Essential Concepts}

We begin by defining concepts that are essential for the causal view of quantum systems.

\subsection{What is a causal influence?}

We assume throughout this section that every (separable) Hilbert space $\ch$ represents a quantum system. Let $U \in \cb(\ch)$ be a unitary operator representing a transformation on the system represented by $\ch$, and let $\cu:=U(\cdot)U^\dagger$ be the corresponding unitary channel. Let $M$ be an arbitrary Hermitian operator associated with the time before $U$ is implemented, and let $N$ be an arbitrary Hermitian operator associated with the time after $U$ is implemented. What does it mean to claim that $M$ influences $N$ through $U$, i.e.\ $M \xrightarrow{U} N$? Our answer is inspired by the quantum causal models of \cite{Allen_2017, barrett2020quantum, barrett2021cyclic, ormrod2023causal}:

\begin{definition} \label{def:influence}
    Given a unitary transformation $U \in \cb(\ch)$, $M \xrightarrow{U} N$ if $[\cu^{-1}(N), M] \neq 0$.
\end{definition}

To get a feel for Definition \ref{def:influence}, assume for the moment that the unitary transformation is the identity, $U=I$. Then $M \xrightarrow{I} N$ if and only if $[N, M]\neq 0$. One interpretation of this expression is that the \textit{generator} represented by $M$ induces a change in the \textit{observable} represented by $N$. (Recall Heisenberg's equation, $\dot N = -\frac{i}{\hbar}[N, M]$.)

More generally, if $U$ is an arbitrary unitary transformation then $M \xrightarrow{U} N$ tells us that $M$ generates a change in $\cu^{-1}(N)$, which is then transformed by $U$ into $N$. In this way, $M$ \textit{indirectly} changes $N$, \textit{via} the unitary $U$. Operationally, if Alice implements the generator $M$ before $U$, and Bob measures the observable $N$ after $U$, then $M \xrightarrow{U} N$ if and only if Alice can signal to Bob (for at least one initial state of the system). 

Although \cite{Allen_2017, barrett2020quantum, barrett2021cyclic, ormrod2023causal} define causal influences between entire systems, Theorem 3.2 of \cite{ormrod2023causal} implies that one system influences another (in the quantum causal models sense) if and only if some Hermitian operator associated with the former influences some Hermitian operator associated with the latter (in the sense of Definition \ref{def:influence}). Hence Definition \ref{def:influence} is a \textit{fine-graining} of the usual notion of influence from quantum causal models. From now on, we will assume that this is the correct approach to causal influence, but see \cite{ormrod2023causal} for a detailed argument. 

\subsection{What is accessibility?} \label{sec:accessibility}

Suppose that the ``system'' $\ch_S$ and ``probe'' $\ch_P$ unitarily interact via $U \in \cb(\ch_S \otimes \ch_P)$. The interaction $U$ is fixed, but an observer can choose the state of the probe before the interaction and perform any measurement she likes on the probe afterwards. Intuitively, $U$ will allow the observer to indirectly access certain observables on $\ch_S$, but not others. But what does it mean exactly for a given observable $M_S$ to be made \textit{accessible} by $U$?

One might be tempted to say that $M_S$ is accessible if it influences at least one observable on the probe, i.e.\ if $\exists N_P: M_S \xrightarrow{U} N_P$. (From now on, we denote this condition using the shorthand $M_S \xrightarrow{U} \ch_P$.) But to see that this definition is misguided, let $\ch_S$ and $\ch_P$ be qubits interacting via the CNOT unitary, defined by ${\rm CNOT}\ket{k}_S \ket{m}_P =\ket{k}_S \ket{m+k}_P$. If the observer prepares the probe in the state $\ket{0}_P$ before the CNOT interaction begins, and measures the probe in the $Z_P$ basis after the interaction is complete, then the outcome probabilities will exactly match the outcome probabilities for a direct measurement of the $Z_S$ basis performed before the interaction. Surely, this means that $Z_S$ is made accessible by the CNOT. Yet $Z_S \not\xrightarrow U \ch_P$, because the \textit{generator} $Z_S$ does not induce any change in probe observables via the CNOT.
Therefore, this definition of accessibility would imply that $Z_S$ is \textit{not} accessible.

For a second attempt at a definition of accessibility, one might say that an observable $M_S$ is accessible if a change in that observable leads, via $U$, to changes in the observables on the probe. One way to formalize this is to say that $M_S$ is accessible if there exists a generator $G_S$ such that $[M_S, G_S]\neq 0$ and $G_S \xrightarrow U \ch_P$. This definition recovers the intuition that $Z_S$ is made accessible by the CNOT interaction,  since e.g.\ $[Z_S, X_S] \neq 0$ and $[{\rm CNOT}^\dagger (I_S \otimes Z_P) {\rm CNOT}, \ X_S \otimes I_P] \neq 0$.

But this is still not the right definition, and again, the CNOT example illustrates why not. Defining the channels $\cc_{\ket{\pm i}}$ on the probe by
\begin{equation}
    \begin{split}
        \cc_{\ket{+ i}} = {\rm Tr}_S{\rm CNOT}(\ket{+i}\bra{+i}_S \otimes (\cdot)) {\rm CNOT}^\dagger \\
        \cc_{\ket{- i}} = {\rm Tr}_S {\rm CNOT}(\ket{-i}\bra{-i}_S \otimes (\cdot)){\rm CNOT}^\dagger,  \\
    \end{split}
\end{equation}
where $\ket{\pm i}:= \frac{1}{\sqrt{2}}(\ket{0} \pm i \ket{1})$,
we can show that $\cc_{\ket{+ i}} = \cc_{\ket{- i}}$,
meaning that one cannot signal to the probe by swapping between the eigenstates of $Y_S$. This makes it intuitively clear that $Y_S$ is not accessible. Yet the very same generator that we just used to argue that $Z_S$ is accessible, namely $X_S$, can be used in the same way to argue that $Y_S$ is accessible.

The moral here is that for $M_S$ to be accessible it is not enough for there to be \textit{one} generator $G_S$ that changes $M_S$ and influences the probe. For $M_S$ to be accessible, the observer should be able to detect \textit{any} change made to $M_S$, regardless of which generator was used to make that change. Formalizing this idea gives us what we claim is the correct definition of accessibility:

\begin{definition} \label{def:accessible}
    Given a unitary interaction $U \in \cb(\ch_S \otimes \ch_P)$, $M_S$ is accessible if for all $G_S$ such that $[M_S, G_S] \neq 0$, $G_S \xrightarrow{U} \ch_P$.
\end{definition}

According to Definition \ref{def:accessible}, $M_S$ is accessible when a change in $M_S$ \textit{implies} an influence on the probe. As the reader can confirm, it follows from Definition \ref{def:accessible} that $Z_S$, but not $Y_S$, is made accessible by the CNOT interaction.

Definition \ref{def:accessible} also yields intuitive verdicts on accessibility in more general scenarios.  For example, a projective-valued measurement (PVM) $\{\pi_S^k\}_k$ is often thought of as being implemented by a Hamiltonian interaction of the form $H=M_S \otimes N_P$, where $M_S$ is an observable with spectral projectors $\{\pi_S^k\}_k$ and $N_P$ is some generator on the probe. This Hamiltonian generates the ``coherent control'' unitary transformations $U(t) = \sum_k \pi_S^k \otimes e^{-im_ktN_P}$, where the $m_k$ are the eigenvalues of $M_S$. As long as $e^{-im_ktN_P}$ is not equal to $ e^{-im_lN_P} $ up to a phase for any $k \neq l$, the full set of observables made accessible by $U(t)$ is indeed the set ${\rm rspan}(\{\pi_S^k\}_k)$ of real linear combinations of PVM elements.

We note one more reassuring consequence of Definition \ref{def:accessible}. Let us call a probe ``useful'' if there is some influence from $\ch_S$ to the probe (i.e.\ there exists some $M_S$ and $N_P$ such that $M_S \xrightarrow U N_P$). Intuition would suggest that a probe is useful if and only if it makes some observables that are nontrivial (i.e.\ not proportional to the identity) accessible. Appendix \ref{app:accessibility_necessary_influence} confirms that Definition \ref{def:accessible} recovers this intuition.

From now on, we take it for granted that Definition \ref{def:accessible} is the right definition of accessibility. With this definition in now hand, one can write down the complete list ${\rm Acc}_S(\ch_P|U)$ of observables made accessible by any given $U$ to an observer who can measure probe $\ch_P$. Such lists will supply us with our causal definition of quantum systems.

\section{Results}

This section formalizes the causal view and derives the algebraic approach to quantum systems.

\subsection{The causal view of quantum systems}

In the previous section, we assumed that Hilbert spaces represented quantum systems. Now we drop that assumption. On the causal view, systems are conceptually downstream of the unitary interactions, in terms of whose causal structure they are defined:

\begin{definition} \label{def:qsystem}
    $\mathscr{S} \subseteq  {\rm Herm}(\cb(\ch_S))$ is a quantum system if there exists some $\ch_P$ and unitary interaction $U \in \cb(\ch_S \otimes \ch_P)$ such that $\mathscr{S}$ is the complete set of accessible observables, $\mathscr{S}= {\rm Acc}_S(\ch_P|U)$.
\end{definition}

That is, a quantum system is any complete list of accessible observables.

Although the causal view does not assume a priori that Hilbert spaces represent systems, it does \textit{derive} that they do from Definition \ref{def:qsystem}. Let $\ch_S$ be any separable Hilbert space, and define the SWAP unitary interaction by ${\rm SWAP}\ket{\psi}_S \ket{\phi}_P= \ket{\phi}_S \ket{\psi}_P$. It is easy to show that the SWAP makes accessible all observables on $\ch_S$, ${\rm Acc}_S(\ch_P|{\rm SWAP}) = {\rm Herm}(\cb(\ch_S))$. Therefore, ${\rm Herm}(\cb(\ch_S))$ is a quantum system. Since the Hilbert space $\ch_S$ is in one-one correspondence with ${\rm Herm}(\cb(\ch_S))$, $\ch_S$ can be taken to represent a system.\footnote{Thus although the causal view does not assume a priori that the $\ch_S$ and $\ch_P$ that appear in the definition of a unitary interaction $U \in \cb(\ch_S \otimes \ch_P)$ represent systems, it does recover this assumption from Definition \ref{def:qsystem}. It thus recovers the idea that interactions are interactions \textit{between} systems, even though interactions are defined independently of systems.}

But more generally, the causal view implies that quantum systems are represented by operator algebras.

\begin{theorem} \label{thm:qsystem}
    For any quantum system $\mathscr{S} \subseteq {\rm Herm}(\cb(\ch_S))$, there exists a unique von Neumann algebra $\ca \subseteq \cb(\ch_S)$ such that $\mathscr{S} = {\rm Herm}(\ca)$.
\end{theorem}

Theorem \ref{thm:qsystem} is proven in Appendix \ref{app:algebra_reps_system}, where $\ca$ is constructed by taking commutants and intersections of von Neumann algebras corresponding to $\ch_S$ and $\ch_P$.

Having shown that every quantum system is represented by a unique von Neumann algebra, one might ask whether every von Neumann algebra defines a quantum system. Appendix \ref{app:algebra_reps_system} shows that this is true at least in the finite-dimensional case by using a generalization of the SWAP interaction.

\subsection{The causal view of classical systems}

If a quantum system is characterized by the possibility of being all that is accessible to a given observer, then what characterizes those quantum systems that behave classically?

According to quantum Darwinism, classical objectivity arises out of the quantum substrate through the spreading of multiple copies of the same piece of information throughout the environment, allowing multiple observers to independently obtain the same knowledge \cite{zurek2003decoherence, zurek2009quantum}. Inspired by this, the causal view defines a \textit{classical quantum system} as any complete list of observables that are \textit{independently accessible} to a pair of possible observers. 

To make this precise, suppose that $\ch_S$ is coupled to \textit{two} probes by a unitary interaction $U \in \cb(\ch_S \otimes \ch_{P_1} \otimes \ch_{P_2})$. We can now apply Definition \ref{def:accessible} of accessibility in two different ways. We can think of $\ch_{P_1}$ as probing $\ch_S \otimes \ch_{P_2}$ and accessing the observables in ${\rm Acc}_{SP_2}(\ch_{P_1}|U)$. Or we can think of $\ch_{P_2}$ as probing $\ch_S \otimes \ch_{P_1}$ and accessing the observables in ${\rm Acc}_{SP_1}(\ch_{P_2}|U)$. Notice that if the same observable is accessible to both probes then it is an observable on $\ch_S$ alone, $M_S \in {\rm Acc}_{S}(\ch_{P_1}, \ch_{P_2}|U) := {\rm Acc}_{SP_1}(\ch_{P_2}|U) \cap {\rm Acc}_{SP_2}(\ch_{P_1}|U)$.

We say that the probes are \textit{causally independent} if $\ch_{P_1} \not\xrightarrow U \ch_{P_2}$ and $\ch_{P_2} \not\xrightarrow U \ch_{P_1}$. An interaction  $U \in \cb(\ch_S \otimes \ch_{P_1} \otimes \ch_{P_2})$ makes $M_S$ \textit{independently accessible} if its probes are causally independent and $M_S \in {\rm Acc}_{S}(\ch_{P_1}, \ch_{P_2}|U)$. According to the causal view, the characteristic feature of classical quantum systems is that they are not only accessible, but independently accessible:

\begin{definition} \label{def:csystem}
    $\mathscr{C}$ is a classical quantum system if there exists a unitary interaction $U \in \cb(\ch_S \otimes \ch_{P_1} \otimes \ch_{P_2})$ with causally independent probes such that $\mathscr{C}$ is the complete set of independently accessible observables, $\mathscr{C} = {\rm Acc}_{S}(\ch_{P_1}, \ch_{P_2}|U)$.
\end{definition}

The causal view does not assume, but rather derives, the link between classicality and commutativity:

\begin{theorem} \label{thm:csystem}
    If $\mathscr{C}$ is a classical quantum system, then there exists a unique commutative operator algebra $\cc$ such that $\mathscr{C}= {\rm Herm}(\cc)$.
\end{theorem}

Theorem \ref{thm:csystem} is proven in Appendix \ref{app:csystem}, which also shows that any finite-dimensional commutative von Neumann algebra represents a classical quantum system.

Theorem \ref{thm:csystem} can be thought of as a ``causal'' or ``Heisenberg picture'' version of the no-cloning theorem \cite{wootters1982single}, forbidding the ``copying'' of noncommuting observables rather than non-orthogonal states.
By the same token, the theorem provides a causal or Heisenberg-picture gloss on quantum Darwinism: classical objectivity emerges from those causal structures that copy and spread information about observables throughout the environment.

Taken together, Theorems \ref{thm:qsystem} and \ref{thm:csystem} comprise our derivation of the algebraic approach to quantum systems from the causal view.

\subsection{An equivalent reformulation of the causal view}

Before moving on to the discussion, we note that the causal view can be equivalently stated in terms of lists of \textit{implementable generators} rather than accessible observables. Given a unitary interaction $U \in \cb(\ch_S \otimes \ch_{P})$, the generator $G_S$ is \textit{implementable} if for every observable $M_S$ such that $[M_S, G_S] \neq 0$, $\ch_{P} \xrightarrow {U} M_S$. Note that, by Definition \ref{def:influence}, causal influences are \textit{reversible} in the sense that $M \xrightarrow U N$ if and only if $N \xrightarrow {U^\dagger} M$. It follows that $U$ makes the \textit{generator} $G_S$ implementable if and only if $U^\dagger$ makes the \textit{observable} $G_S$ accessible.

Therefore, Definition \ref{def:qsystem} is logically equivalent to a definition of a quantum system as any full list of implementable generators. Similarly, Definition \ref{def:csystem} is logically equivalent to a definition of a classical quantum system as any full list of \textit{independently implementable} generators. Theorems \ref{thm:qsystem} and \ref{thm:csystem} can also be derived from these alternative definitions.

\section{Discussion:  What is a system?} \label{sec:conc}

On the epistemic view, a system can be thought of as a list of Hermitian operators ``closed under deduction'': the list contains any observable whose expectation can always be deduced from the expectations of all others on the list. On the causal view,  a system can be thought of as a list of Hermitian operators ``closed under causation'', containing any Hermitian operator made accessible (or implementable) by all causal structures that make accessible (or implementable) all others on the list. 

$\mathscr{R}$ is the simplest example of a list of Hermitian operators on whose systemhood the two views disagree. According to the epistemic view, $\mathscr{R}$ is a system because there is a possible observer, namely Polly from above, who knows all and only the expectation values of observables in $\mathscr{R}$. But according to the causal view, $\mathscr{R}$ is not a system, because any unitary interaction $U$ that allows Polly to access every observable in $\mathscr{R}$ must also allow her access to observables outside $\mathscr{R}$, e.g.\ $Y$. (This follows from Theorem \ref{thm:qsystem} and the fact that the smallest algebra containing $\mathscr{R}$ is the full algebra of operators on the qubit.) The reason that closure under products is relevant to systemhood is that any observable obtained from complex sums of products of accessible observables is accessible.

While the algebraic approach follows from the causal view, it is inconsistent with the epistemic view, since the latter ascribes systemhood to $\mathscr{R}$. Appendix \ref{app:modifed} shows that two natural modifications to the epistemic view also fail to recover the algebraic approach. If, as tradition would have it, all quantum systems are represented by operator algebras, then the epistemic view should be rejected in favour of the causal view.

But it is not only the weight of tradition that backs up the causal view: it is also supported by the perspective that physics is essentially about how information flows through physical interactions, rather than how deductions are made in the mind of an observer. In this regard, it is worth emphasizing that although we appealed to the notion of an observer to pedagogically motivate Definition \ref{def:accessible} of accessibility, that definition makes no mention of observers, and ultimately can be understood simply as a notion of information about an observable on $\ch_S$ flowing through $U$ into $\ch_P$. 

Our two main results respectively suggest directions for further research into (1) the fundamental quantum domain and (2) the emergence from it of the classical domain. Theorem \ref{thm:qsystem} shows that the structure of quantum systems as traditionally understood is shaped by causal influence, and raises the question of whether the concept of causation might be used to derive more aspects of quantum theory, or perhaps even the theory as a whole. Theorem \ref{thm:csystem} raises the possibility of a ``purely causal'' approach to quantum Darwinism, and, more generally, of a purely causal approach to decoherence and the emergence of classicality -- one that might take quantum state, which makes no appearance in Definition \ref{def:influence}, entirely out of the picture.

\acknowledgements It is a pleasure to thank the organizers and participants of the 2025 New Directions in the Foundations of Physics conference for interesting discussions that led to this work, and likewise to thank Roberto D. Baldij\~ao, Tom Galley, Simon Langenscheidt, David Schmid, and Augustin Vanrietvelde for helpful discussions and comments on a draft. NO was supported by Perimeter Institute for Theoretical Physics. Research at Perimeter Institute is supported in part by the Government of Canada through the Department of Innovation, Science and Economic Development and by the Province of Ontario through the Ministry of Colleges and Universities.

\appendix

\section{Can a modified epistemic view recover the algebraic approach?} \label{app:modifed}

In this paper, we have pointed out that the epistemic view of quantum systems is incompatible with the algebraic approach.  But could a modified version of the epistemic view---one that preserves the core idea that a system is a list of observables ``closed under deduction'', while formalizing it differently---fare better? This appendix does not seek to answer this question definitively, but it does show that two obvious modifications to the epistemic view fail to recover the algebraic approach.

In the introduction, we noted that an observer who knows the expectation value of each observable in $\mathscr{R}$ cannot \textit{in general} deduce the expectation values of observables outside of $\mathscr{R}$. However, she can do in some finely-tuned special cases. For example, if she finds that $\langle Z\rangle_\rho=+1$, then she can deduce the expectation values of all of the other observables on the qubit. Motivated by this example, let us consider a modification to the epistemic view that regards a set $\mathscr{S}$ of observables as a system if and only if there is \textit{no} possible complete set of expectation values of all observables in $\mathscr{S}$ from which the expectation value of any observable outside of $\mathscr{S}$ can be inferred.

On this modified version of the epistemic view, $\mathscr{R}$ is not a system, since, as we have just seen, the expectation values of observables outside of $\mathscr{R}$ can, in certain cases, be inferred from the expectation value of $Z \in \mathscr{R}$. But by the same token, this modified epistemic view denies systemhood to $\mathscr{Z}:={\rm rspan}(I, Z)$. On the other hand, the algebraic approach must regard $\mathscr{Z}$ as a system, since $\mathscr{Z}= {\rm Herm}(\cz)$, where $\cz$ denotes the von Neumann algebra of operators diagonal in the $Z$ basis. More generally, this modified epistemic view denies systemhood to any complete set of observables contained in a non-factor algebra $\mathscr{S}= \bigoplus_i \cb(\ch_i)$ (where there is more than one term in the direct sum), i.e.\ to any superselected quantum system. Therefore, this version of the epistemic view fails to recover the algebraic approach.

Let us now motivate a different modification to the epistemic view. In the introduction, we pointed out that an observer who knows the expectation value of each observable in $\mathscr{R}$ cannot \textit{deduce} the exact expectation values of observables outside of the set. However, she can make nontrivial inferences about them. For example, she could use her knowledge of $\langle X \rangle_\rho$ and $\langle Z \rangle_\rho$ together with the inequality
\begin{equation}
     \langle Y \rangle^2_\rho  \quad \leq \quad 1 - \langle X \rangle^2_\rho - \langle Z \rangle^2_\rho
\end{equation}
to place precise and, in most cases, nontrivial numerical bounds on $\langle Y \rangle_\rho$. 

Let us therefore consider a modification to the epistemic view that regards $\mathscr{S}$ as a system if and only if knowledge of the expectation values of observables in $\mathscr{S}$ does not permit any nontrivial inferences about the expectation values of observables outside of $\mathscr{S}$. 

Natural though it may seem at first glance, the implications of this modification of the epistemic view are disastrous. Like the previous modification, it must deny that non-factor algebras represent quantum systems. But it also must deny that even factor algebras of dimension greater than 1 represent quantum systems, or, equivalently, it must deny that \textit{Hilbert spaces} of dimension greater than 1 represent quantum systems. For example, consider a pair of qubits $\ch_A$ and $\ch_B$ and define the Bell state $\ket{\Phi^+} := \frac{1}{\sqrt{2}}(\ket{0}_A \ket{0}_B + \ket 1_A \ket{1}_B)$. If $\langle Z_A \rangle_{\rho_A}=+1$, then $\langle \ket{\Phi^+}\bra{\Phi^+} \rangle_{\rho_{AB}} \leq \frac{1}{2}$, but $\ket{\Phi^+}\bra{\Phi^+} \not\in {\rm Herm}(\cb(\ch_A))$. Therefore, this modified epistemic view implies that ${\rm Herm}(\cb(\ch_{A}))$ is not a system.

Both of the proposed modifications to the epistemic view are based on a stricter interpretation of the general idea behind the epistemic view: that a system is a set of observables that is somehow closed under the inferences an observer can make from them. Both backfire for the same reason: at least on a traditional understanding of systems, knowledge of observables on a system \textit{does} facilitate certain inferences about other observables.

\section{Proofs of novel results} \label{app:proofs}

This appendix proves a number of statements and theorems from the main text. The full derivation of the algebraic approach from the causal view is found in Sections \ref{app:algebra_reps_system} and \ref{app:csystem}.

\subsection{Probes are useful if and only if they make nontrivial observables accessible} \label{app:accessibility_necessary_influence}

Here, we prove a claim from Section \ref{sec:accessibility} about the ``usefulness'' of probes. We say that $G_S$ \textit{helps} make an observable $M_S$ accessible if $M_S$ is accessible and $[M_S, G_S] \neq 0$. We then have the following lemma:

\begin{lemma} \label{lemma:helps}
    Given a unitary interaction $U \in \cb(\ch_S \otimes \ch_P)$ and operator $G_S \in {\rm Herm}(\cb(\ch_S))$, $G_S \xrightarrow U \ch_{P}$ if and only if $G_S$ helps make at least one observable $M_S$ accessible. 
\end{lemma}

\textit{Proof of Lemma \ref{lemma:helps}}. The ``if'' direction follows immediately from the definitions of helping and accessibility. For the ``only if'' direction, we will rely on the first three of the standard results about operator algebras mentioned in the proof of Theorem \ref{thm:qsystem} and proven in Appendix \ref{app:algebra}. (We therefore recommend that dedicated readers study the proof of Theorem \ref{thm:qsystem} before this one.) An observable $M_S$ is accessible if and only if $M_S \otimes I_P\in {\rm Herm}(\cs \cap {\rm Herm}(\cu^{-1}(\mathcal{P}))')' = (\cs \cap \cu^{-1}(\mathcal{P})')'$.
Now suppose that $G_S$ does not help make any $M_S$ accessible. This means that $G_S$ commutes with all accessible observables, $G_S \otimes I_P \in (\cs \cap \cu^{-1}(\mathcal{P})')''.$ By the double commutant theorem for von Neumann algebras, this last statement is equivalent to $G_S \otimes I_P \in \cs \cap \cu^{-1}(\mathcal{P})'$, i.e.\  $G_S \not\xrightarrow U \ch_P$. \textit{End of proof.} \\

If a probe is useful, then there exists some $G_S$ such that $G_S \xrightarrow U \ch_P$. By Lemma \ref{lemma:helps}, this $G_S$ must help make some observable $M_S$ accessible, and this $M_S$ must be nontrivial since $[M_S, G_S] \neq 0$. Conversely, suppose that there exists some nontrivial accessible observable $M_S$. Nontriviality implies that there exists some generator $G_S$ such that $[M_S, G_S] \neq 0$. Accessibility then implies that, for this $G_S$, $G_S \xrightarrow U \ch_P$. Therefore, the probe is useful.

\subsection{Proofs on quantum systems} \label{app:algebra_reps_system}

\textit{Proof of Theorem \ref{thm:qsystem}.} 
Appendix \ref{app:algebra} gives a formal definition of a von Neumann algebra and proves four standard results which are simply stated in the rest of this paragraph. Firstly, any von Neumann algebra $\ca \subseteq \cb(\ch)$ is equal to the set of complex linear combinations of its Hermitian elements, $\ca={\rm cspan}({\rm Herm}(\ca))$. By the bilinearity of the commutator, it follows that $\ca' = {\rm Herm}(\ca)'$, where $\ca'$ is the \textit{commutant} defined $\ca':=\{M\in \cb(\ch)|[A, M]=0 \quad \forall A\in \ca\}$. Secondly, the intersection of any pair of von Neumann algebras is a von Neumann algebra. Thirdly, if $\ca$ is a von Neumann algebra, then $\ca'$ is also a von Neumann algebra. Finally, if $\tilde \ca :=  \{M_A \otimes I_B| M_A \in \ca\}\subseteq \cb(\ch_A\otimes \ch_B)$ is a von Neumann algebra, then the set $\ca \subseteq \cb(\ch_A)$ is also a von Neumann algebra.

If $\mathscr{S}$ is a system, then there exists some unitary interaction $U \in \cb(\ch_S \otimes \ch_P)$ such that $\mathscr{S}={\rm Acc}_S(\ch_P|U)$. Let $\cs:= \cb(\ch_S) \otimes I_P$ be the von Neumann algebra of bounded operators that act locally on $\ch_S$, and similarly $\mathcal{P}:= I_S \otimes \cb(\ch_P)$. The generators on $\ch_S$ that do not influence $\ch_P$ are those contained in the set ${\rm Herm}(\cs \cap {\rm Herm}(\cu^{-1}(\mathcal{P}))')$. By the contrapositive form of Definition \ref{def:accessible}, $\mathscr{S} = {\rm Herm}({\rm \ca})$, where $\ca \subseteq \cb(\ch_S)$ is defined by
\begin{equation}
    \begin{split}
        \ca \otimes I_P=& \cs \cap 
{\rm Herm}((\cs \cap {\rm Herm}(\cu^{-1}(\mathcal{P}))')' \\
        =& \cs \cap (\cs \cap \cu^{-1}(\mathcal{P})'))'
    \end{split}
\end{equation}
By the second equality, $\ca \otimes I_P$ is obtained by commutants and intersections of von Neumann algebras, and is therefore itself a von Neumann algebra. Therefore, $\ca$ is a von Neumann algebra. To show that $\ca$ is the unique algebra satisfying $\mathscr{S} = {\rm Herm}({\rm \ca})$, we take the complex span of each side to obtain ${\rm cspan}(\mathscr{S}) = \ca$. \textit{End of proof.} \\

Theorem \ref{thm:qsystem} states that (given Definition \ref{def:qsystem}) every quantum system is represented by a von Neumann algebra. We now prove the converse, assuming for technical simplicity that $\ca \subseteq \cb(\ch_S)$ is an algebra of operators on \textit{finite-dimensional} Hilbert space $\ch_S$. In that case, $\ca$ has the form $\ca= \bigoplus_{i=1}^n I_{S_L^i} \otimes \cb(\ch_{S_R}^i)$ for some decomposition of the Hilbert space $\ch_S = \bigoplus_{i=1}^n \ch_{S_L}^i \otimes \ch_{S_R}^i$ (see Theorem III.1.1 of \cite{davidson1996c}). For the moment, assume that each $\ch_{S_R}^i$ has the same dimension, $d$. Let $\ch_P$ have dimension $d$, and let $U = \bigoplus_{i=1}^n I_{S_L^i} \otimes {\rm SWAP}_{S_R^iP}$. Then ${\rm Acc}_S(\ch_P|U) = {\rm Herm}(\ca)$, meaning (by Theorem \ref{thm:qsystem})  that the algebra $\ca$ uniquely represents the system ${\rm Acc}_S(\ch_P|U)$. 

Now let us relax our earlier assumption and allow the different $\ch_{S_R}^i$ to have different dimensions. Let $d$ be the greatest dimension among these, and let $\ch_P$ have dimension $d$. For each $i$, write $\ch_P = \ch_{P^{i0}} \oplus \ch_{P^{i1}}$ for some subspace $\ch_{P^{i0}}$ satisfying ${\rm dim}(\ch_{S_R^i})={\rm dim}(\ch_{P^{i0}})$. Define $U^i := I_{S_L^i} \otimes ({\rm SWAP}_{S_R^iP^{i0}} \oplus U_{S_R^iP^{i1}})$ for an arbitrary unitary operator $U_{S_R^iP^{i1}}$, and let $U := \bigoplus_{i=1}^nU^i$. Again, ${\rm Acc}_S(\ch_P|U) = {\rm Herm}(\ca)$.

\subsection{Proofs on classical quantum systems} \label{app:csystem}

\textit{Proof of Theorem \ref{thm:csystem}.} If $\mathscr{C}$ is a classical quantum system $\cs$, then there exists some unitary interaction $U \in \cb(\ch_S \otimes \ch_{P_1} \otimes \ch_{P_2})$ with causally independent probes such that $\mathscr{C} = {\rm Acc}_{S}(\ch_{P_1}, \ch_{P_2}|U)$. By an argument similar to the proof of Theorem \ref{thm:qsystem}, $\mathscr{C} = {\rm Herm}(\cc$), where $\cc$ is the von Neumann algebra defined by
\begin{equation} \label{eq:commutative_algebra} 
\begin{split} 
        \tilde \cc  =& \Big( \cs \mathcal{P}_2 \cap (\cs \mathcal{P}_2 \cap \cu^{-1}(\mathcal{P}_1)')'\Big) \\ & \quad  \cap  \Big( \cs \mathcal{P}_1 \cap (\cs \mathcal{P}_1 \cap \cu^{-1}(\mathcal{P}_2)')' \Big)   \\
        =& \cs \cap (\cs \mathcal{P}_2 \cap \cu^{-1}(\mathcal{P}_1)')' \cap (\cs \mathcal{P}_1 \cap \cu^{-1}(\mathcal{P}_2)')',
\end{split}
\end{equation}
where $\tilde \cc := \cc \otimes I_{P_1} \otimes I_{P_2}$ and $\cs \mathcal{P}_2$ is the algebra $I_{P_1} \otimes \cb(\ch_S \otimes \ch_{P_2})$, etc. As in the proof of Theorem \ref{thm:qsystem}, the uniqueness of $\cc$ follows from taking the complex span of each side of $\mathscr{C} = {\rm Herm}(\cc$). 

To complete the proof, we will now show that $\cc$ is commutative. Note that $\cc$ is commutative if and only if $\tilde \cc$ is commutative. By Eq.\ \ref{eq:commutative_algebra}, 
\begin{equation} \label{eq:friend2}
    \tilde \cc \subseteq (\cs\mathcal{P}_2 \cap \cu^{-1}(\mathcal{P}_1)')'.
\end{equation}
We will now show that also 
\begin{equation} \label{eq:friend1}
    \tilde \cc \subseteq \cs \mathcal{P}_2 \cap \cu^{-1}(\mathcal{P}_1)', 
\end{equation} 
from which the commutativity of $\tilde \cc$ (and thus $\cc$) will follow. Causal independence implies that $\ch_{P_2} \not\xrightarrow U \ch_{P_1}$, or, equivalently, that  $\cu^{-1}(\mathcal{P}_1) \subseteq \cs \mathcal{P}_1 = \mathcal{P}_2'$. Obviously, $\cu^{-1}(\mathcal{P}_1) \subseteq \cu^{-1}(\mathcal{P}_2)'$. Combining these last two inclusion relations gives us $\cu^{-1}(\mathcal{P}_1) \subseteq \cs \mathcal{P}_1 \cap \cu^{-1}(\mathcal{P}_2)'$. Taking the commutant of both sides gives $(\cs\mathcal{P}_1 \cap \cu^{-1}(\mathcal{P}_2)')' \subseteq \cu^{-1}(\mathcal{P}_1)'$. But by Eq.\ \ref{eq:commutative_algebra}, $\tilde \cc \subseteq (\cs\mathcal{P}_1 \cap \cu^{-1}(\mathcal{P}_2)')'$. By the transitivity of inclusion, $\tilde \cc \subseteq \cu^{-1}(\mathcal{P}_1)'$.
By Eq.\ \ref{eq:commutative_algebra}, $\tilde \cc \subseteq \cs \mathcal{P}_2$. Combining these last two inclusion relations gives Eq.\ \ref{eq:friend1}. \textit{End of proof.} \\

We note that the proof of Theorem \ref{thm:csystem} only relies on one half of the causal independence assumption, namely $\ch_{P_2} \not\xrightarrow U \ch_{P_1}$. If the definition of independent accessibility were weakened to allow causal influence in only one direction between the probes, then Theorem \ref{thm:csystem} would still hold. 

We now show that any commutative algebra qualifies as a classical quantum system, assuming for technical simplicity that $\ch_S$ is finite-dimensional.  Any commutative algebra has the form $\cc={\rm rspan }(\{\pi_{S}^k\}_{k=1}^n)$  for some PVM $\{\pi_{S}^k\}_{k=1}^n$. Define $U := \sum_{k=1}^n \pi_S^k \otimes V_{P_1}^{(k)} \otimes V_{P_2}^{(k)}$ where each $\{V_{P_m}^{(k)}\}_k$ is a collection of unitaries satisfying $V_{P_m}^{(k)} \not\propto V_{P_m}^{(k')}$ for all $k \neq k'$. Then ${\rm Acc}_{S}(\ch_{P_1}, \ch_{P_2}|U) = {\rm Herm}(\cc)$, meaning that ${\rm Herm}(\cc)$ is a classical quantum system.

\section{Standard results on operator algebras} \label{app:algebra}


For completeness, this appendix gives a formal definition of a von Neumann algebra and derives four standard results about operator algebras used in the proof of Theorem \ref{thm:qsystem} and throughout Appendix \ref{app:proofs}. The interested reader can learn more about operator algebras from various textbooks, e.g.\ \cite{davidson1996c}.

Unless one ignores the infinite-dimensional case, the definition of a von Neumann algebra requires the notion of a set $\cs \subseteq \cb(\ch)$ being \textit{weakly closed.} Recall that a sequence of complex numbers, $z_n$ for $n \in \mathbb{N}$, \textit{converges} to a limit $z$ if for every $\epsilon>0$, there exists some $N \in \mathbb{N}$ such that $|z_n - z| < \epsilon$ for all $n \geq N$. In that case, $z$ is called the \textit{limit} of the sequence. A sequence of operators, $M_n$ for $n \in \mathbb{N}$, \textit{weakly converges} to $M$ if for every $\ket{\psi}, \ket{\phi} \in \ch$, the sequence of complex numbers $\bra{\phi} M_n \ket{\psi}$ converges to $\bra{\phi} M \ket{\psi}$. In that case, $M$ is called the \textit{weak limit} of the sequence of operators. A set $\cs \subseteq \cb(\ch)$ of operators is \textit{weakly closed} if it contains the weak limit of every weakly convergent sequence it contains.

\begin{definition} \label{def:algebra}
    A set of bounded operators $\cs \subseteq \cb(\ch)$ is a \textit{von Neumann algebra} if it contains the identity operator on $\ch$ and all adjoints, complex linear combinations, and products of its members, and is weakly closed. 
\end{definition}

In the finite-dimensional case, weak closure follows from the other conditions in Definition \ref{def:algebra}, and can thus be omitted.

The first standard result on operator algebras assumed in the proof of Theorem \ref{thm:qsystem} was that any von Neumann algebra $\ca$ is obtained by taking complex combinations of its Hermitian elements, $\ca = {\rm cspan}({\rm Herm}(\ca))$. This is an immediate corollary of the following lemma.
\begin{lemma} \label{lemma:spanherm}
    If a bounded set $\cs \subseteq\cb(\ch)$ of operators contains all adjoints and complex linear combinations of its members, then $\cs = {\rm cspan}({\rm Herm}(\cs))$.
\end{lemma}

\textit{Proof of Lemma \ref{lemma:spanherm}.} Let $\cs \subseteq\cb(\ch)$ contain all adjoints and complex linear combinations of its members. For any $S \in \cs$, the operators $A:= \frac{1}{2}(S^\dag + S)$ and $B:= \frac{i}{2}(S^\dag - S)$ are each contained in $\cs$. $S$ can be written as a complex linear combination $S = A + iB$ of these operators. \textit{End of proof.} \\

It is clear from Definition \ref{def:algebra} that the intersection $\ca \cap \cb$ of any pair of von Neumann algebra $\ca$ and $\cb$ is a von Neumann algebra. This was the second standard result assumed in the proof of Theorem \ref{thm:qsystem}.

The third standard result was that the commutant of any von Neumann algebra is itself a von Neumann algebra. This is a corollary of the following lemma.

 \begin{lemma} \label{lemma:commutant}
    If a bounded set $\cs \subseteq\cb(\ch)$ of operators contains the adjoint of each of its members, then $\cs'$ is a von Neumann algebra.
\end{lemma}

\textit{Proof of Lemma \ref{lemma:commutant}}. Let $\cs \subseteq \cb(\ch)$ be an arbitrary set of bounded operators, and let $\cs'$ be its commutant. Since the identity operator commutes with all operators, it is contained in $\cs'$. Since the commutator is bilinear and $\cs'$ is defined as the set commuting with $\cs$, $\cs'$ contains arbitrary complex linear combinations of its members. By the identity $[S, MN]= [S, M]N + M[S, N]$, $\cs'$ contains arbitrary products of its members.

We now show that for any set of bounded operators $\cs \subseteq \cb(\ch)$, $\cs'$ is weakly closed. Let the operators $M_n$ for $n \in \mathbb{N}$ be a sequence contained in $\cs'$ with limit $M$. For any $S \in \cs$, $[S, M_n]=0$ for all $n$, meaning that the sequences $\bra{\phi} S M_n \ket{\psi}$ and $\bra{\phi} M_n S \ket{\psi}$ are the same, and their limits, $\bra{\phi} S M \ket{\psi}$ and $\bra{\phi} M S \ket{\psi}$, are equal. Since this holds for all $\ket{\psi}, \ket{\phi} \in \ch$, it follows that $SM=MS$. Since this in turn holds for all $S \in \cs$, it follows that $M \in \cs'$.

Finally, suppose that $\cs$ contains the adjoint of each of its members. If $M \in \cs'$, then for any $S\in \cs$, $[S, M^\dagger] = [S, M^\dagger]^{\dagger \dagger}= [M, S^\dagger]^\dagger=0$. Hence $M^\dagger \in \cs'$. It follows that $\cs'$ contains the adjoint of each of its members. Combining this with the results of previous paragraphs, it follows that $\cs'$ is von Neumann algebra. \textit{End of Proof.} \\

The fourth and final standard result used in the proof of Theorem \ref{thm:qsystem} was that, for any tensor product Hilbert space $\ch_A \otimes \ch_B$ and set $\ca \subseteq \cb(\ch_A)$, if $\tilde  \ca := \ca \otimes I_B$ is a von Neumann algebra then so too is $\ca$. If $\tilde \ca$ is a von Neumann algebra, it is straightforward to show that $\ca$ contains the identity on $\ch_A$ and is closed under complex linear combinations, products, and adjoints. In the finite-dimensional case, this completes the proof, but more generally one still has to show that $\ca$ is weakly closed. To do so, we make use of the following lemma, which is perhaps easier to believe than to rigorously prove:
\begin{lemma} \label{lemma:convergence}
    If the sequence $M_A^{(n)}$ weakly converges to $M_A$, then the sequence $M_A^{(n)} \otimes I_B$ weakly converges to $M_A \otimes I_B$. 
\end{lemma}

Once this lemma is accepted, the remainder of the argument is straightforward. Suppose that the sequence $M_A^{(n)}$ is contained in $\ca$. It follows that the sequence $M_A^{(n)} \otimes I_B$ is contained in $\tilde \ca$. If $M_A^{(n)}$ weakly converges to $M_A$ then Lemma \ref{lemma:convergence} tells us that $M_A^{(n)} \otimes I_B$ weakly converges to $M_A \otimes I_B$, and then the weak closure of $\tilde \ca$ implies that $M_A \otimes I_B \in \tilde \ca$, which in turn implies that $M_A \in \ca$. Thus $\ca$ is weakly closed. \\

\textit{Proof of Lemma \ref{lemma:convergence}}. The proof will proceed in three stages. First, we will show that for any weakly convergent sequence $M_A^{(n)}$, the sequence $M_A^{(n)} \otimes I_B$ weakly converges when restricted to the subspace $\cd \subseteq \ch_A \otimes \ch_B$ of vectors that can be written as a finite sum of tensor product vectors, $\ket{\psi}= \sum_{k=1}^N \ket{\psi_k}_A \ket{\psi_k'}_B$. Second, we will show that the sequence $M_A^{(n)} \otimes I_B$ is \textit{uniformly bounded} (as defined below). Third, we will use the \textit{denseness} (defined below) of $\cd$ and the uniform bound on $M_A^{(n)} \otimes I_B$ to show that the sequence is weakly convergent on the full Hilbert space $\ch_A \otimes \ch_B$. 

\textit{Stage 1: convergence on $\cd$}. Let $M_A^{(n)}$ be a sequence of operators weakly converging to $M_A$. The reader can confirm that if sequences $x_n$ and $y_n$ converge to $x$ and $y$ respectively, then the sequence $z_n:=ax_n + by_n$ converges to $ax+by$. It follows that for any $\ket{\psi}, \ket{\phi} \in \cd$, the sequence
\begin{equation}
\begin{split}
    z_n :=&  \bra{\phi} M_A^{(n)} \otimes I_B \ket{\psi} \\  =& \sum_{k=1}^N \sum_{m=1}^M \bra{\phi_m} M_A^{(n)}\ket{\psi_k}_A \braket{\phi'_m|\psi_k'}_B
\end{split}
\end{equation}
converges to $z:= \sum_{k=1}^N \sum_{m=1}^M \bra{\phi_m} M_A\ket{\psi_k}_A \braket{\phi'_m|\psi_k'}_B = \bra{\phi} M_A \otimes I_B \ket{\psi}$. In other words, $M_A^{(n)} \otimes I_B$ weakly converges to $M_A \otimes I_B$ when restricted to the subspace $\cd$.

\textit{Stage 2: uniform boundedness.} A sequence of complex numbers $w_n$ is \textit{uniformly bounded} if there exists some finite positive number $p$ such that $|w_n| \leq p$ for all $n$. 
Similarly, a sequence of operators $M_n$ is uniformly bounded if there exists some finite positive number $p$ such that $||M_n|| \leq p$, where $||M_n||$ denotes the operator norm of $M_n$.

We now show that any converging sequence $w_n$ of complex numbers is uniformly bounded. If $w_n$ converges to $w$, then there exists some $N$ such that, for all $n \geq N$, $|w_n - w|\leq 1$.  By the triangle inequality for complex numbers, it follows that $|w_n| \leq 1+ |w|$ for all $n \geq N$. This implies that the sequence $w_n$ is uniformly bounded (either by $p:= 1 + |w|$, or by the maximum value of $|w_n|$ for $n <N$). 

Since $M_A^{(n)}$ weakly converges, it follows that, for any $\ket{\psi}, \ket{\phi} \in \ch_A $, the sequence $\bra{\phi}M_A^{(n)} \ket{\psi}$ is uniformly bounded. 
It then follows by the uniform boundedness principle (also known as the Banach-Steinhaus theorem) that the sequence $M_A^{(n)}$ itself is uniformly bounded by some positive number $C$. Finally, since  $||M_A^{(n)} \otimes I_B|| = ||M_A^{(n)}|| \ ||I_B|| = ||M_A^{(n)}||$, the sequence $M_A^{(n)} \otimes I_B$ has the same uniform bound $C$.

\textit{Stage 3: extension to} $\ch_A \otimes \ch_B$. This stage makes use of the concept of a dense subspace. A sequence of states $\ket{\psi^{(k)}}$ \textit{converges} to the limit $\ket{\psi}$ if the sequence of complex numbers $|| \ket{\psi^{(k)}}-\ket{\psi}||$ converges to 0, where $||\ket{\varphi}||:= \sqrt{\braket{\varphi|\varphi}}$. A subspace of a Hilbert space is \textit{dense} if the set of all limits of its converging sequences is the full Hilbert space. By definition of the tensor product of Hilbert spaces, $\cd$ is a dense subspace of $\ch_A \otimes \ch_B$. 

Our goal is to show that for any $\ket{\psi}, \ket{\phi} \in \ch_A \otimes \ch_B$ and $\epsilon >0$, we can find an $N$ such that $|\bra{\phi} M_A^{(n)} \otimes I_B \ket{\psi} - \bra{\phi} M_A \otimes I_B \ket{\psi}| < \epsilon$ for all $n \geq N$. To that end, let $\ket{\psi^{(k)}}$, $\ket{\phi^{(k)}} \in \cd$ be sequences of vectors respectively converging to $\ket{\psi}$ and $\ket{\phi}$. Define the shorthand $M_n:= M_A^{(n)} \otimes I_B$ and $M := M_A \otimes I_B$. To obtain an upper bound on $|\bra{\phi} M_n \ket{\psi} - \bra{\phi} M \ket{\psi}|$ , we can write \small
\begin{equation} 
\begin{split}
        \bra{\phi} M_n \ket{\psi} - \bra{\phi} M \ket{\psi} 
&= \Big( \bra{\phi} M_n \ket{\psi} - \bra{\phi^{(k)}} M_n \ket{\psi^{(k)}} \Big) \\
&\quad + \Big( \bra{\phi^{(k)}} M_n \ket{\psi^{(k)}} - \bra{\phi^{(k)}} M \ket{\psi^{(k)}} \Big) \\
&\quad + \Big( \bra{\phi^{(k)}} M \ket{\psi^{(k)}} - \bra{\phi} M \ket{\psi} \Big)
\end{split}
\end{equation}
\normalsize and then apply the triangle inequality to find that \small
\begin{equation} \label{eq:three_terms}
\begin{split}
        \big| \bra{\phi} M_n \ket{\psi} - \bra{\phi} M \ket{\psi} \big|
& \leq  \big|\bra{\phi} M_n \ket{\psi} - \bra{\phi^{(k)}} M_n \ket{\psi^{(k)}}\big|\\
&\quad + \big| \bra{\phi^{(k)}} M_n \ket{\psi^{(k)}} - \bra{\phi^{(k)}} M \ket{\psi^{(k)}} \big| \\
&\quad + \big| \bra{\phi^{(k)}} M \ket{\psi^{(k)}} - \bra{\phi} M \ket{\psi} \big|.
\end{split}
\end{equation}
\normalsize 
Another application of the triangle inequality gives us an upper bound for the first term on the right hand side of Eq.\ \ref{eq:three_terms},
\begin{equation} \small \hspace{-0.8cm}
\begin{split}
\big|\bra{\phi}& M_n \ket{\psi} - \bra{\phi^{(k)}} M_n \ket{\psi^{(k)}}\big|
\\ &\leq \big|\bra{\phi - \phi^{(k)}} M_n \ket{\psi}\big|
    + \big|\bra{\phi^{(k)}} M_n \ket{\psi - \psi^{(k)}}\big| \\
&\leq \|\ket{\phi - \phi^{(k)}}\| \, \|M_n\| \, \|\ket \psi\|
    + \|\ket {\phi^{(k)}}\| \, \|M_n\| \, \|\ket{\psi - \psi^{(k)}}\| \\
&\leq C \big( \|\ket{\phi - \phi^{(k)}}\| \, \|\ket{\psi}\| 
              + \|\ket{\phi^{(k)}}\| \, \|\ket{\psi - \psi^{(k)}}\| \big),
\end{split}
\end{equation}  \normalsize
where $\ket{\psi - \psi^{(k)}}:= \ket{\psi} - \ket{\psi^{(k)}}$, etc. Notice how the uniform boundedness of $M_n$ has allowed us to obtain an upper bound that is independent of $n$. Due to the convergence of the sequences $\ket{\psi^{(k)}}$, $\ket{\phi^{(k)}}$, this upper bound can be made arbitrarily small by choosing a large enough $k$. Similarly, one can obtain an arbitrarily small bound for the third term on the right of Eq.\ \ref{eq:three_terms} by choosing a large enough $k$. 

To complete the argument, consider some arbitrary $\epsilon>0$. Choose a large enough $k$ that the sum of the first and third terms on the right of Eq.\ \ref{eq:three_terms} is at most $\epsilon'$, where $\epsilon' <\epsilon$. Since $M_n$ converges when restricted to $\cd$, the second term converges. Therefore, there exists some $N$ such that for all $n \geq N$ the second term is less than $\epsilon-\epsilon'$. It follows that $|\bra{\phi} M_n \ket{\psi} - \bra{\phi} M \ket{\psi}| < \epsilon$. \textit{End of proof.} \\

\vspace{0.5cm}

\bibliography{refs}

\end{document}